# A simple approach for vortex core visualization

Jiajia Li and Pablo M. Carrica*

IIHR-Hydroscience and Engineering, The University of Iowa, Iowa City, IA 52242, USA

*Corresponding author, pablo-carrica@uiowa.edu

## SUMMARY

We propose a method to visualize vortex cores based on manipulation of the pressure field produced by isolated vortices in incompressible flow. Under ideal conditions the function $D = 2|\nabla p|/\nabla^2 p$ yields an approximate distance to vortex centerlines. As opposed to local methods to identify coherent structures, isosurfaces of *D* produce a field of vortex tubes equidistant to the vortex core center which, ideally, are independent of vortex intensity or size. In contrast to other line-vortex identification methods, which typically rely on algorithms to detect vortex core lines and frequently need complex implementations, the proposed method can be computed from the local Eulerian velocity and pressure fields as easily as vortex identification methods such as the *Q* and $\lambda_2$ criteria. $D = 2|\nabla p|/\nabla^2 p$ results in the exact distance to the core center for a Rankine vortex and is in general valid for the region of a vortex where there is pure rotation, yielding an approximation to the distance farther from the core in other simple one-dimensional vortex models. The methodology performs well in all tests we attempted, though limitations are presented and discussed. The method is demonstrated for a canonical Burgers vortex, a Bodewadt vortex, homogeneous isotropic turbulent flow, the wake of a propeller, a heaving plate, a turning containership and the airwake of a surface combatant. The proposed method helps to better visualize vortical flow fields by displaying vortex cores, complementing methods like *Q* and $\lambda_2$ which display vortical volumes.

KEY WORDS: Vortices, Computational Fluid Dynamics.

## 1. Introduction

While the definition of a vortex is still being debated, the need to visualize flow fields has led to the development of a considerable number of vortex identification methods. Though none of these methods is perfect or universal, due to limitations and a natural consequence of the lack of a vortex definition, many of them have been in use for a number of years, and are frequently tested by users to find the method that best fits their particular problem and needs.

A vortex filament is generally recognized as a region where rigid body rotation has a maximum and constitutes the center of the vortex. The definition of a vortex core is more complicated since it involves identifying the boundary between the rotational and the irrotational flow, becoming subjective (see different points of view on this in [1-3]). This definition of the interface between rotational and irrotational flow is already arbitrary for simple one-dimensional vortex considering the diffusion of vorticity by viscosity, and becomes even more so for complex three dimensional flows where vortices may not be symmetric, and are subject to interaction with other vortices and flow features like pressure gradients and strain.

In recent review papers, Epps [4] and Volkov *et al.* [5] broadly classify vortex identification methods into region-type methods, which attempt to identify the volume of the vortices, and line-type methods, which seek to detect the lines joining the vortex centers. These last methods are typically expensive and require specific algorithmic implementations. Vortex identification methods are also classified as Eulerian, where Eulerian quantities (velocity and/or pressure fields) are used to identify vortices, or Lagrangian, typically based on particle trajectories. In addition, Eulerian methods can be local, depending only on local variables, or global, which require larger regions in the domain. Eulerian, local methods are usually the easiest to implement and are therefore most widely used. We present in this paper an Eulerian local method to identify vortex cores, which can be easily combined with Eulerian local vortex identification methods to provide the location of vortex cores.

## 1.1. *Eulerian vortex identification methods*

The *Q*-criterion vortex identification method was one of the first Eulerian local approaches, and we dwell here in some detail as it is relevant to the discussion in section §2. It was proposed by Hunt *et al.* [6], defining vortex regions where *Q*, the second invariant of velocity gradient $\nabla v$, is positive. The method enjoyed immediate popularity due to its simplicity and ability to efficiently display vortical structures. While the *Q*-criterion has some shortcomings and alternative methods have been proposed since (either Eulerian or Lagrangian [1, 3, 7, 8]), it remains widely used today. One of the criticisms to the *Q*-criterion is that, though is Galilean invariant, is not an objective method [2] and produces spurious structures when the system of reference is rotating. *Q* is defined as

$$2Q = \|\mathbf{\Omega}\|^2 - \|\mathbf{S}\|^2 \qquad (1)$$

where $\mathbf{\Omega} = (\nabla v - \nabla v^\mathrm{T})/2$ and $\mathbf{S} = (\nabla v + \nabla v^\mathrm{T})/2$ are the antisymmetric and symmetric components representing rates of rotation and strain. A relation between *Q* and the Laplacian of the pressure can be obtained by taking the divergence of the dimensionless incompressible Navier-Stokes equation,

$$\frac{\partial v}{\partial t} + (v \cdot \nabla) v = -\nabla P + \frac{1}{\mathrm{Re}} \nabla^2 v \qquad (2)$$

where reference values of length and velocity are used to non-dimensionalize Eq. (2). The result is [9]

$$Q = \frac{1}{2} \nabla^2 P \qquad (3)$$

It is important to note that, in addition to the second invariant of the velocity gradient, *Q* can be replaced by the Laplacian of the pressure, and that a positive *Q* can be interpreted as a negative source for the Poisson Eq. (3), $\nabla^2 P - 2Q = 0$. A positive *Q* naturally results in a local pressure drop due to a negative pressure source. Thus the *Q*-criterion to define a vortex can be also stated as regions where sources in the pressure Poisson equation are negative. In the case of a rigid body rotation this source is

$$\nabla^2 P - 2\Omega^2 = 0 \qquad (4)$$

We note here that, while vortices always cause negative sources for the pressure Poisson Eq. (3), other sources can be present in the flow that significantly affect the pressure field caused by vortices, thus the pressure minimum will not necessarily be located in the vortex core.

A second criticism to the *Q*-criterion is related to the need to stablish a threshold, such that any point with $Q > Q_t$ is identified as a vortex if a pressure minimum occurs within this region (see discussion in [9]). In complex flows with a wide range of spatial scales a low $Q_t$ will produce false positives, and overly busy structures. A high $Q_t$, on the other hand, will display less and tighter structures, clearing the picture and allowing easier observation of the stronger vortices, but weaker vortices are lost when they may be relevant.

Other Eulerian local vortex identification methods have been proposed with various degrees of success. There is no consensus on what method is ultimately superior; instead all have strengths and weaknesses. Jeong & Hussain [3] proposed the $\lambda_2$-criterion, which uses a negative second eigenvalue of the second order tensor $\mathbf{\Omega}^2 + \mathbf{S}^2$ to define a vortex. The $\lambda_2$-criterion, as the *Q*-criterion, is widely used and relatively simple to evaluate, though computation of the eigenvalues of $\mathbf{\Omega}^2 + \mathbf{S}^2$ and ordering is more involved than the computation of *Q*, which only requires derivatives of the velocity. Chakraborty *et al.* [1] introduced the $\lambda_{ci}/\lambda_{cr}$ enhanced swirling strength criterion, an improvement to the original $\lambda_{ci}$ swirling strength criterion method of Zhou *et al.* [10].

## 1.2. *Vortex core line identification methods*

Vortex core line identification methods are mostly based on local or global algorithms that detect the center of rotation of vortices. One of the most widely used methods was developed by Sujudi & Haimes [11], and involves detection of a center of rotation in individual cells. The velocity field is interpolated on each tetrahedra and $\nabla v$ is computed. If the eigenvalues of $\nabla v$ has one real $\lambda_R$ and two complex conjugate $\lambda_C$ eigenvalues, then the velocity field is projected into the plane normal to the eigenvector corresponding to $\lambda_R$. The process to find the center of rotation involves detecting if the line of rotation intersects the tetrahedron faces, if two points are found then the center line crosses the element. This algorithm is simple and easy to parallelize, though it tends to fail for vortex core lines with high curvature [12]. This problem was studied and solutions proposed by Roth [13] in terms of parallel vector operators, which can be used to express most definitions of vortex cores. By choosing appropriate pairs of vectors and Boolean operators, Roth [13] was able to group the methods of Sujudi & Haimes [11], Levy *et al.* [14], Banks & Singer [15], Kida & Miura [16] and Strawn *et al.* [17] into a single algorithm implementation framework.

To our knowledge, all current vortex core line detection methods fall in the category of algorithmic methods, with different degrees of difficulty. Most of the proposed schemes can be very complex to implement and rarely used. The method to visualize vortex core proposed in this paper, though it cannot find the exact center of the vortex, can be used to approximate the core lines using isosurfaces of small *D* as shown in subsequent sections. In addition, it is Eulerian and does not require specific implementations, and can be used with the capabilities to process velocity and pressure fields available in most commercial or open source software to postprocess flow solutions.

## 2. Distance to a vortex core line

In this section we present a methodology to produce vortex tubes at a given distance around vortex center lines as isosurfaces of a function *D* obtained from the Eulerian local velocity and pressure.

### 2.1. *The distance function D*

In absence of strong stretching by external strain, the inner portion of the core of a vortex can be approximated by rigid body rotation, thus the azimuthal velocity is a linear function with the radial distance to the vortex center line and the angular rotation velocity in an inertial frame of reference, $v_\theta = \Omega r$. The pressure gradient is in this case

$$\frac{dP}{dr} = \Omega^2 r \qquad (5)$$

Equation (5) is exact near the core line for isolated axisymmetric vortices, and an approximation when the vortices are immersed in pressure gradients generated by other flow features like nearby vortices or geometrical boundaries. Since according to Eq. (4) the Laplacian of the pressure in this case is $\nabla^2 P = 2\Omega^2$, dividing by the Laplacian of the pressure we obtain a function $D$ that provides the distance to the vortex center line

$$D = r = \frac{dP}{dr} \bigg/ \left[ \frac{1}{2r} \frac{d(r dP/dr)}{dr} \right] \qquad (6)$$

In a general three dimensional flow $D$ can be expressed as

$$D = \frac{2|\nabla P \cdot \mathbf{n}|}{\nabla \cdot (\nabla P \cdot \mathbf{n}) \mathbf{n}} \qquad (7)$$

where $(\nabla P \cdot \mathbf{n})\mathbf{n}$ is the pressure gradient projected in the direction of a unit vector $\mathbf{n}$ normal to the vortex axis. $\mathbf{n}$ can be obtained from the eigenvector corresponding to $\lambda_2$ as defined in the $\lambda_2$ criterion [3]. If the pressure gradient is dominated by the pressure field produced by the vortex, which as will be discussed in §3 is in most situations the case since $D$ is relevant in locations inside vortex cores, we can then simplify Eq. (7) to

$$D = \frac{|\nabla P|}{\frac{1}{2}\nabla^2 P} \qquad (8)$$

since in this case the pressure gradient and Laplacian are dominated by the local vortex. Though Eq. (8) can be computed if the pressure field is available from either CFD or experiments, it involves second derivatives of the pressure, and in some cases may result in unacceptable noise. Recalling that the Laplacian of the pressure and $Q$ are related by Eq. (3), then a version of Eq. (8) that requires only first derivatives of velocity and pressure and results in smoother contours is

$$D = \frac{|\nabla P|}{Q} \qquad (9)$$

Finally, since only regions where $Q > 0$ are of interest, then the distance to the vortex core line can be written as

$$D = \frac{|\nabla P|}{\max(Q, \epsilon)} \qquad (10)$$

where $\epsilon$ is a small positive number to prevent dividing by zero and make $D$ large when $Q$ is small. If strong pressure gradients not caused by rotation are present, due for instance to local flow acceleration, then Eq. (7) can be simplified to

$$D = \frac{|\nabla P \cdot \mathbf{n}|}{\max(Q, \epsilon)} \qquad (11)$$

which in some cases yields better results than Eq. (10), at the cost of a more complex computation. Alternatively, $\lambda_2$ can be used to compute $D$, since $Q = -\lambda_2/2$ in planar flows

$$D = \frac{|\nabla P \cdot \mathbf{n}|}{\max(-\lambda_2/2, \epsilon)} \qquad (12)$$

which essentially computes $Q$ in the plane of rotation, removing effects of pressure variations in direction of the axis of rotation. Eq. (12), however, is not immune to external pressure fields in the plane of rotation.

### 2.2. Simple vortex models

Gerz *et al.* [18] presents several one-dimensional vortex models in the context of a study on aircraft wing tip vortices. All models share two important properties. Since the center of the core rotates as a rigid body, the vorticity gradient normal to the vortex center line is zero at the vortex axis, $\partial\omega/\partial r = 0$, implying that the azimuthal velocity increases linearly with $r$, $v_\theta \sim r$. The second is that the vorticity vanishes far from the core, and thus the azimuthal velocity decreases as $v_\theta \sim 1/r$. The simplest vortex model is the Rankine vortex, where the vorticity is constant within the vortex core, and zero outside,

$$v_\theta = \begin{cases} r & r \leq 1 \\ 1/r & r > 1 \end{cases} \quad (13)$$

where the velocity is nondimensionalized with $\Gamma/(2\pi a)$, the azimuthal velocity at a characteristic vortex core size $a$ as defined in Leweke et al. [19]. $a$ is used as reference length for the equations in dimensionless form. The Rankine vortex suffers from a discontinuity in the velocity gradient at the vortex core edge $r = 1$. A more realistic model is the Lamb-Oseen vortex model,

$$v_\theta = 1/r\left(1 - e^{r^2}\right) \quad (14)$$

Using the one-dimensional version of *D*, Eq. (6), we obtain in cylindrical coordinates

$$D = \frac{dP}{dr} \bigg/ \left[\frac{1}{2r}\frac{d(rdP/dr)}{dr}\right] = v_\theta \bigg/ \frac{dv_\theta}{dr} \quad (15)$$

For the Rankine vortex, *D* is then

$$D = \begin{cases} +r & r \leq 1 \\ -r & r > 1 \end{cases} \quad (16)$$

and for the Lamb-Oseen vortex is

$$D = \frac{r\left(1 - e^{-r^2}\right)}{e^{-r^2}\left(2r^2 + 1\right) - 1} \quad (17)$$

The azimuthal velocity $v_\theta$ and the function *D* for Rankine and Lamb-Oseen vortices are shown in Fig. 1. Note that the core for the Lamb-Oseen vortex is not located exactly at $r = 1$; some authors apply a correction on the exponent to displace the core back to $r = 1$ [18], but that results in a deviation from the rigid body rotational speed for $r \to 0$.

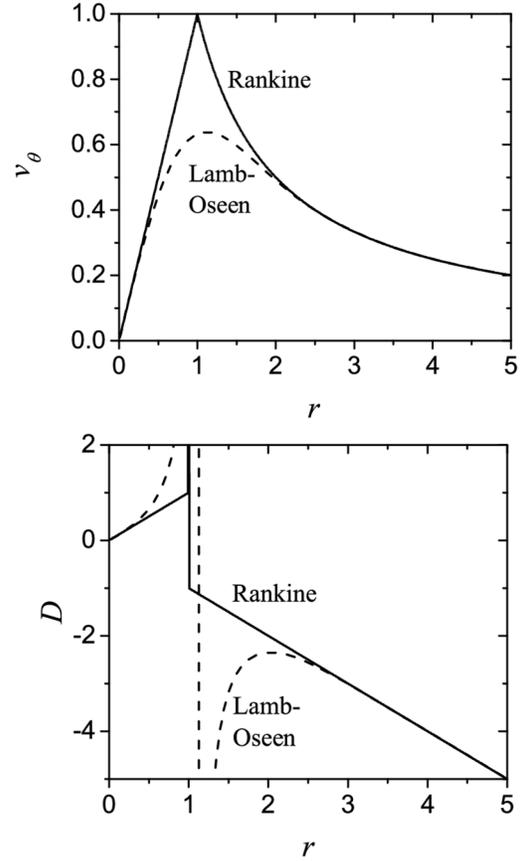

Fig. 1: Rankine and Lamb-Oseen vortices, $v_\theta$ (top) and D (bottom).

Since derived for the rigid body region of the vortex core, *D* is only valid for small *r*, which occurs when *Q* is positive. Notice that for the Lamb-Oseen vortex *D* is approximately 10% larger than *r* for $r = 0.3$ (30% of the vortex core), providing an estimate of the error incurred when the *D* function is displayed. Adding straining to the Lamb-Oseen vortex results in the Burgers vortex model, a good example of a vortex affected by an external pressure gradient

$$v_r = -\beta r, v_\theta = 1/r\left[1 - e^{-r^2}\right], v_z = 2\beta z \quad (18)$$

In Eq. (18) $\beta = 4\pi\nu/\Gamma$ is a dimensionless parameter that controls the strength of the straining velocity field $(v_r, v_z)$ respect to the strength of the vortical structure represented by $v_\theta$. The resulting total dimensionless pressure gradient is shown in Fig. 2 for different $\beta$ and $z$ values. For weak straining, $\beta = 0.05$, the external pressure gradient is negligible compared to the pressure gradient induced by the flow rotation, resulting in no deformation of the predicted tubes using the approximate Eq. (10) respect to predictions with the more accurate Eq. (7) (see Fig. 3). A stronger straining flow with $\beta = 0.2$ causes a high pressure gradient in the axial direction of the vortex $(z)$, resulting in higher total pressure gradient as $|z|$ increases. The result is a distortion of the isosurface of *D* further out from the straining plane at higher $|z|$, deviating from the desired vortex tube. Notice that using only the pressure gradient normal to the axis, Eq. (7), the vortex tube is recovered with little effect on the distance to the center, as shown in Fig. 3. This is explained by the little effect that the straining field has in the radial pressure gradient, depicted in Fig. 2.

$$\frac{\partial P}{\partial r} = \frac{-v_r^2}{r} + \frac{v_\theta^2}{r}, \quad \frac{\partial P}{\partial \theta} = 0, \quad \frac{\partial P}{\partial z} = \frac{v_z^2}{z} \quad (19)$$

In the Burgers vortex the radial inflow produces a pressure gradient caused by the straining velocity field, independent of the vortical structure represented by $v_\theta$, resulting in a simple example where a pressure field, external to the vortex, affects the behaviour of different methodologies to compute *D*. It must be noted that *Q* is not affected by the straining velocity field, with $Q = 0$ isosurfaces yielding a cylinder of radius $R = 1.1a$. In the case of low to moderate strain, the pressure gradient is dominated by the azimuthal component of the velocity and isosurfaces of $D = 0.4a$ from Eq. (10) produce a tube with actual distance to the vortex axis of approximately $R = 0.355a$, indistinguishable from results using Eq. (7) (see Fig. 2, top). For $\beta = 0.05$ the difference between the total pressure gradient and the pressure gradient due to the azimuthal velocity is 0.03 at $r = 0$, and quickly decreases to less than 0.01 for larger $r$. A stronger straining field causes a considerable pressure field, which causes Eq. (10) to predict a deformed distance function *D*, while Eq. (7) still produces good results (see Fig. 2, bottom). In most practical cases use of Eq. (7) produces results that are very similar to those produced by Eq. (10). While more accurate and immune to effects of external pressure fields on the ability to display vortex tubes, Eq. (7) will in most cases result in noisier images caused by the computation of the second derivatives of pressure.

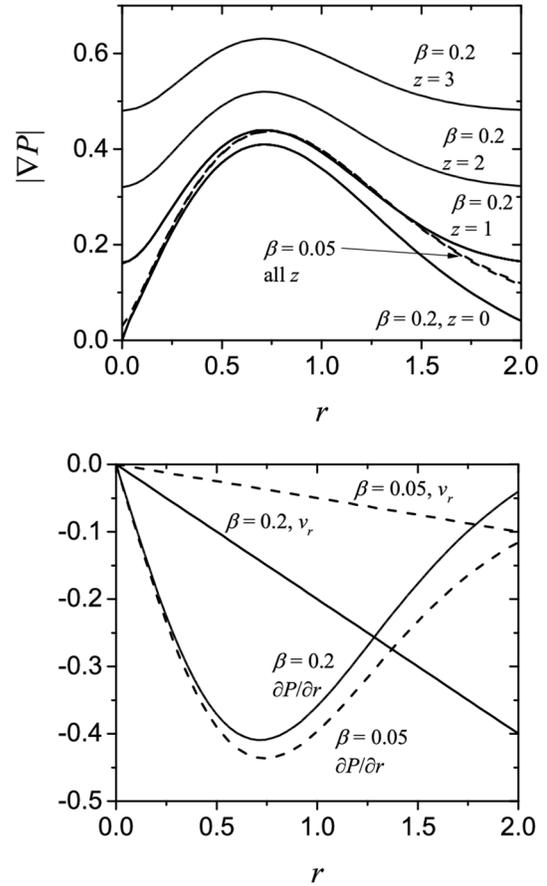

Fig. 2: Burgers vortex: total pressure gradient (top) and radial pressure gradient and velocity (bottom) for various values of $\beta$ and $z$.

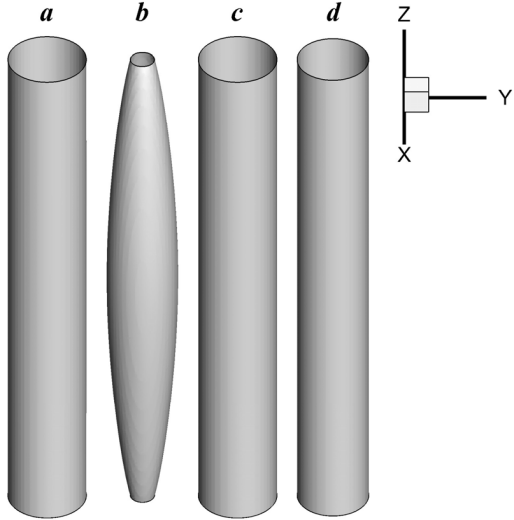

Fig. 3: Isosurfaces of $D = 0.5$ for $\beta = 0.05$ (*a, c*) and $\beta = 0.2$ (*b, d*) with $D$ computed using Eq. (10) (a, b) and Eq. (7) (c, d).

## 3. Discussion and examples

In this and following sections $D$ is computed using Eq. (10) unless otherwise stated, selecting $D$ with values of a few grid points to show the vortex cores. Smaller values of $D$ make the isosurfaces look like vortex core lines, but values smaller than the grid spacing cannot display isosurfaces. Note that, since small values of $D$ maintain the displayed tubes inside the rigid rotation vortex cores, coloring the isosurfaces of $D$ with absolute vorticity provides an approximation to the vorticity in the core.

### 3.1. *Limitations*

The method proposed relies in detecting the rigid rotating core of a vortex based on the assumption that the local pressure gradient is caused by isolated vortices. This assumption of course fails when significant pressure gradients caused by other sources are present. One example is the Burgers vortex shown in §2.2, where strain causes an external pressure gradient that results in a deformation of the tubes of constant $D$. We show in this section that for a wide range of flows the proposed method reveals vortex cores well, but presence of walls may cause strong pressure gradients that mask the pressure gradients resulting from weaker vortices. External pressure gradients on a vortex caused by nearby vortices are in general small and do not cause significant effects on $D$. This is because the pressure gradient induced by a vortex in the irrotational external region decays as $dP/dr \sim 1/r^3$. For a pair of Rankine vortices $\mathcal{V}_1$ and $\mathcal{V}_2$ at a distance $b$ larger than the characteristic radius of either vortex, direct superposition of individual pressure gradients in the core of vortex $\mathcal{V}_1$ with characteristic radius $a_1$ results in

$$\frac{dP/dr}{dP_1/dr} \cong 1 + \frac{\Gamma_2^2}{\Gamma_1^2} \frac{1}{(b/a_1 - 1)^3} \qquad (20)$$

where $\mathcal{V}_2$ is a nearby vortex. Equation. (20) does not use the complex Navier-Stokes solutions to the interaction between the two vortices [19], but provides a first-order estimation of the pressure effects that nearby vortices cause on a local vortex. If a smaller vortex $\mathcal{V}_2$ is located inside the core of a co-rotating local larger vortex $\mathcal{V}_1$, in the core of the smaller vortex $D$ can be estimated as

$$D \cong \frac{\Omega_1^2 r + \Omega_2^2 (b-r)}{\Omega_1^2 + \Omega_2^2} \qquad (21)$$

since $Q \cong \Omega_1^2 + \Omega_2^2$. Since smaller vortices tend to spin faster than larger vortices, Eq. (21) implies that $D$ favors smaller vortices, just as $Q$ and $\lambda_2$ do. The preceding discussion highlights two limitations of the proposed method. The first is that the computation of $D$ is affected by external pressure gradients (not caused by the local vortex), resulting in deformation of the tubes displayed by isosurfaces of $D$. Results show that strong external pressure gradients can mask weaker vortices, but pressure gradients caused by nearby vortices are typically innocuous. A second limitation is that larger vortices tend to be masked by smaller vortices, a trend also observed in most other Eulerian vortex identification methods [20]. A third limitation is that the proposed method is restricted to incompressible flows as shown in its derivation. Performance of *D* when applied to non-symmetric vortices has not been studied, but the example in §3.4 seems to show that elliptic vortices can be displayed by isosurfaces of *D*.

### 3.2. Bodewadt vortex

Bodewadt [21] studied a vortex of rotational rate $\Omega$ normal to a solid wall and extending to infinity in the opposite direction. Schwiderski & Lugt [22] presented solutiuons of the Bodewadt vortex and the similar problem of the von Karman vortex, questioning the validity of the original Bodewadt solution as it is unable to predict boundary layer separation at the axis of rotation. We use the original solution of Bodewadt [21], as done by Jeong & Hussain [3], because while simpler still reveals limitations of vortex identification methods. Defining reference length $\sqrt{\nu/\Omega}$ and velocity $\sqrt{\Omega\nu}$, the solution can be expressed in a cylindrical coordinate system as

$$\begin{aligned} u_r &= rF(z) \\ u_\theta &= rG(z) \\ u_z &= H(z) \\ P &= \frac{1}{2}r^2 + P_0(z) \end{aligned} \quad (22)$$

where $F(z)$, $G(z)$, $H(z)$ are determined by the following expressions

$$\begin{aligned} F^2 - G^2 + HF' &= -1 + F'' \\ 2FG + HG' &= G \\ H'' - HH' &= P_0' \\ 2F + H' &= 0 \end{aligned} \quad (23)$$

with boundary conditions

$$\begin{aligned} F(0) &= 0, \, G(0) = 0, \, H(0) = 0 \\ F(\infty) &= 0, \, G(\infty) = 1 \end{aligned} \quad (24)$$

The Bodewadt vortex develops a boundary layer that reduces the rotational velocity of the vortex near the wall, inducing an inflow toward the vortex axis, as shown in Fig. 4. The flow is similar to the "tea cup" flow, and vortex identification methods should display a vortex reaching all the way to the wall. Jeong & Hussain [3] showed that the *Q*-criterion fails to identify the whole volume as a vortex. Figure 4 shows that $\lambda_2$ also fails to identify the near-wall region as a vortex, as local strain dominates at the boundary layer. Notice that Jeong & Hussain [3] claim that $\lambda_2$ remains negative all the way to the wall, but observation of the solution, also shown in their Fig. 11, reveals a small positive zone near the wall. Isosurfaces of *Q* or $\lambda_2$, however, are unable to provide any useful information, since the vortex is present everywhere and thus the volume should be unbounded. Isosurfaces of *D* = 0.1 computed from Eq. (7) can display the vortex axis almost to the wall, while coloring with vorticity shows the effect of the wall on the vortex strength. When *D* is computed with Eq. (12), it fails to display the core close to the wall as $\lambda_2$ is positive, but still displays the location of the core further out. This example nicely shows how isosurfaces of *D* can complement isosurfaces of *Q* or $-\lambda_2$ to display vortex structures and cores.

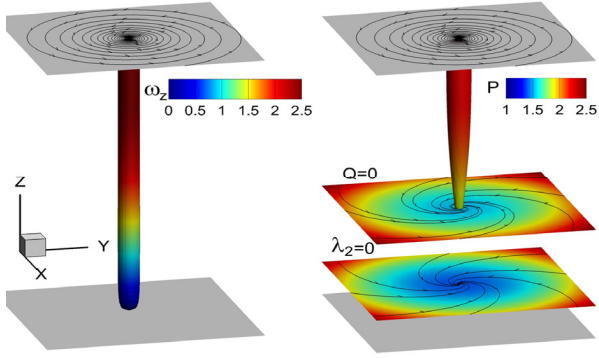

Fig. 4: Bodewadt vortex. Isosurface of $D = 0.1$ computed with Eq. (7) colored with axial vorticity (left). Isosurfaces of $D = 0.1$ computed with Eq. (12) colored with pressure (right). The right panel also shows isosurfaces of $Q = 0$ and $\lambda_2 = 0$, with streamlines over those surfaces.

### 3.3. *Homogeneous isotropic turbulence*

This example uses an instantaneous solution of the homogeneous isotropic turbulence DNS simulation by Li et al. [23], which was obtained in a $1024^3$ uniform grid in a domain extending $[0, 2\pi]$ in all three directions, for $\text{Re}_\lambda = 418$. Results are displayed for a fraction of the original solution at time step $t = 10$, covering half of the grid in each direction (size $512^3$). Figure 5 shows isosurfaces of $D = 0.012$ (top), approximately two cells in size, and $Q = 500$ (bottom), colored with pressure. Notice the presence of large-scale pressure fluctuations, of scale much larger than the resolved vortices shown by either $Q$ or $D$. These large-scale fluctuations do not seem to affect the display of the vortex tubes of size $D = 0.012$, as the large-scale structure pressure gradients are mild. In regions of weaker vortices where $Q = 500$ does not display structures, as the corner near ($x_{max}$, $y_{min}$, $z_{min}$), $D$ produces vortex tubes of approximately equal size as desired. Stronger vortices, which are characterized by low pressure in the core, are displayed clearly as blue tubes in Fig. 5 (top). The difference with isosurfaces of $Q$ is more dramatic for larger vortices, where low values of $D$ locate the isosurfaces deep into the core where the pressure is lower, while the Q-criterion produces larger structures naturally located farther from the vortex axis where pressure is higher. While weaker vortices in regions of higher pressure are mostly absent for $Q = 500$, they are displayed well by $D$. This is clear near the corner ($x_{max}$, $y_{max}$, $z_{min}$) and around the center of the $y_{max}$ face. On the other hand, $Q$ shows the size of the vortices and, by not showing weaker vortices, complements $D$ as a vortex visualization tool. The turbulent field shown in Fig. 5 contains too much information and can hardly be used for visualization. Filtering the velocity and pressure fields reveals larger scales, not discernible when using the original flow field. For example, the pressure field can be filtered by

$$\bar{P}(\mathbf{x}) = \int_{-\infty}^{\infty} P(\mathbf{x}) G(\mathbf{x} - \mathbf{r}) d\mathbf{r} \qquad (25)$$

where $G$ is the Gaussian filter operator. $\bar{P}$ is then a low-pass filtered pressure. The smaller pressure scales can be then obtained as the deviation $P' = P - \bar{P}$. A similar procedure can produce a velocity deviation $\mathbf{u}'$, and $\mathbf{u}'$ and $P'$ can be used to compute $D$ from Eq. (10). We employ a Gaussian operator with standard deviation of 10 grids cells to isolate larger structures. Figure 6 shows isosurfaces of $Q = 20$ for the filtered solution along with isosurfaces of $D = 0.0075$ for the deviation, removing structures with pressures higher than $P = -0.5$ for the smaller scales to clear the picture. The resulting larger vortical structures displayed by isosurfaces of $Q = 20$ exhibit lower rotation rates than smaller structures, and higher pressures. Smaller scale structures displayed by $D$ tend to rotate around the larger structures, but are also present inside the isosurfaces of $Q = 20$, displaying locally the lowest pressures in the field.

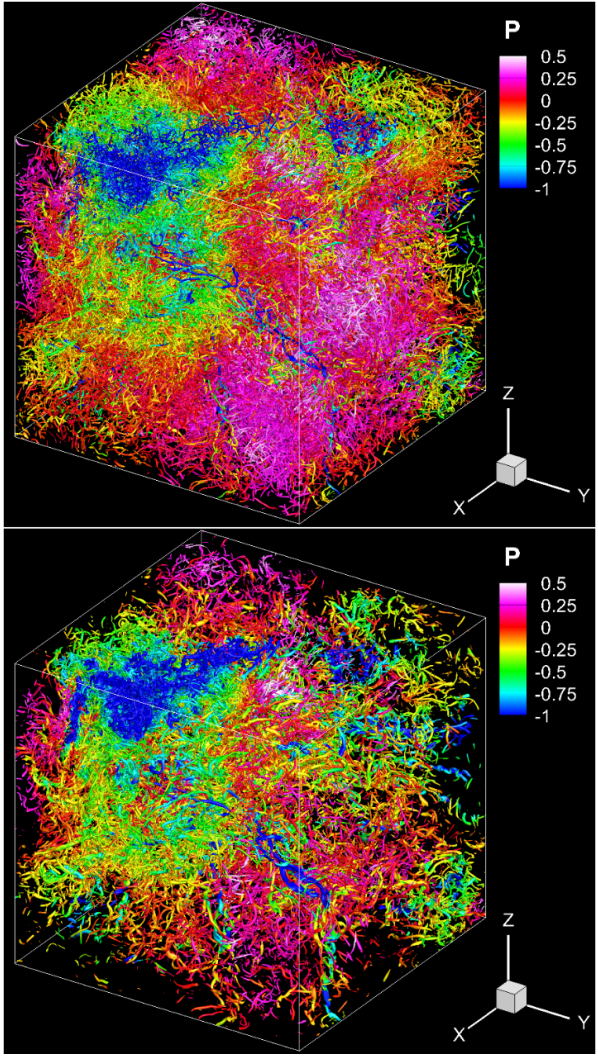

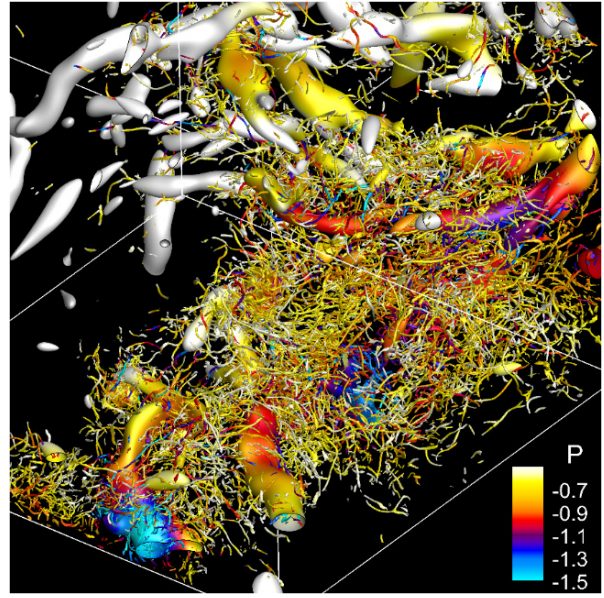

Fig. 6: Homogeneous isotropic turbulence in a periodic box. Isosurfaces of $Q = 20$ of the filtered velocity field and isorufaces of $D = 0.0075$ of the deviation velocity and pressure fields, coloured with pressure. The vortex cores displayed with $D$ are subject to a cutoff showing only pressures lower than $P = -0.5$.

Fig. 5: Homogeneous isotropic turbulence in a periodic box. Isosurfaces of $D = 0.012$ (top) and $Q = 500$ (bottom) coloured with pressure.

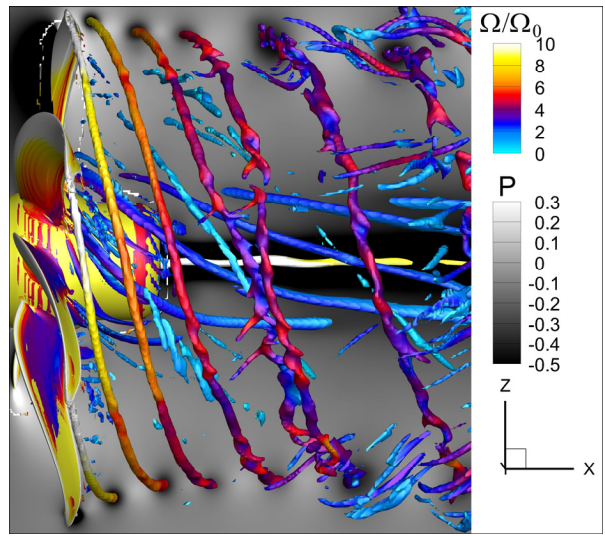

Fig. 7: Propeller E1619 operating at $J = 0.3$. Isosurfaces of $D = 0.0125D_0$ colored with local rotation rate $\Omega/\Omega_0$.

### 3.4. *Propeller wake instabilities*

E1619 is a 7-bladed generic submarine propeller designed at INSEAN (now INM) in Rome, Italy. The propeller has been used as a benchmark to test CFD capabilities to predict propeller performance and wakes [24]. At low advance coefficients the tip vortices exhibit elliptic instabilities, merging and eventual breakdown. Some of these effects have been discussed by Di Mascio, A. *et al.* [25]. In this section we show example results for an advance coefficient $J = V_0 / (\Omega_0 D_0) = 0.3$, where $V_0$ is the advance speed, and $\Omega_0$ and $D_0$ are the propeller rotational speed and diameter, respectively. Figure 7 shows isosurfaces of $D = 0.0125 D_0$ colored with local rotation rate $\Omega = \|\mathbf{\Omega}\|$ nondimensionalized with the propeller rotational speed, and pressure at the centerplane. In a heavily loaded propeller with large number of blades the pitch between tip vortices is small compared to the propeller diameter and vortex radius, resulting in strong vortex interaction. The rotation rates of the tip vortices when shedding exceeds $10\Omega_0$, decaying as they evolve in the propeller's wake. The blade tip vortices are co-rotating helical vortices that develop elliptic instabilities [19], as discussed by Leweke *et al.* [26] for a horizontal wind turbine rotor. These vortex instabilities are characterized by short-wavelength perturbation on the vortex core shape, and can be clearly seen in the tip vortices in Fig. 7 as they develop after the vortices are shed from the blade tip. Merging of two continuous tip vortices can also be observed in Fig. 7 (fifth and sixth vortices from the left), as the elliptic instability continuously growing into the propeller wake, generating a large number of streamwise vortices as the tip vortices breakdown. The very strong hub vortex and the weaker blade root vortices are also clearly exposed by isosurfaces of *D*.

### 3.5. *Heaving plate*

A heaving plate produces a rich vortical system, with periodic tip vortex separation and trailing vortices that form a reverse von Karman streets [27]. In this case a simulation was performed with a zero-thickness square plate of side *L* with free stream velocity *V*, at a Reynolds number $\text{Re} = VL/\nu = 2000$. The plate was subject to a sinusoidal heave oscillation with amplitude $a/L = 0.25$ and a period $TV/L = 1$. A fully structured grid with 18 M grid points was used. Figure 8 shows instantaneous vortical structures displayed with isosurfaces of $Q = 60$ and $D = 0.015$. Isosurfaces of *D* show fairly constant diameter, indicating that *D* is showing an approximate constant distance to the vortex centerlines while no artificial vortex cores are observed. This example, however, displays well regions where *D*, and also *Q*, fail to display vortices. Vortex interaction does not affect *D* significantly for vortices not too close to each other. In regions of weaker and vanishing vorticity, as displayed by isosurfaces of *D* colored with vorticity magnitude, isosurfaces of *D* tend to deviate from the distance to the core center. This is a natural consequence of low pressure gradients caused by the weaker vortices, becoming comparable to pressure gradients caused by the primary flow of the moving plate or by nearby stronger vortices. The vortex cores predicted by *D* can be compared with the vortex cores predicted by the algorithm of Sujudi and Haimes [11] in the bottom of Fig. 8, where it can be noticed that this particular vortex core line identification method based on velocity tensor eigenvalues fails to predict weaker vortices. Notice that while Sujudi and Haimes method predicts a vortex line, in Fig. 8 is shown as a tube with volume to facilitate visualization.

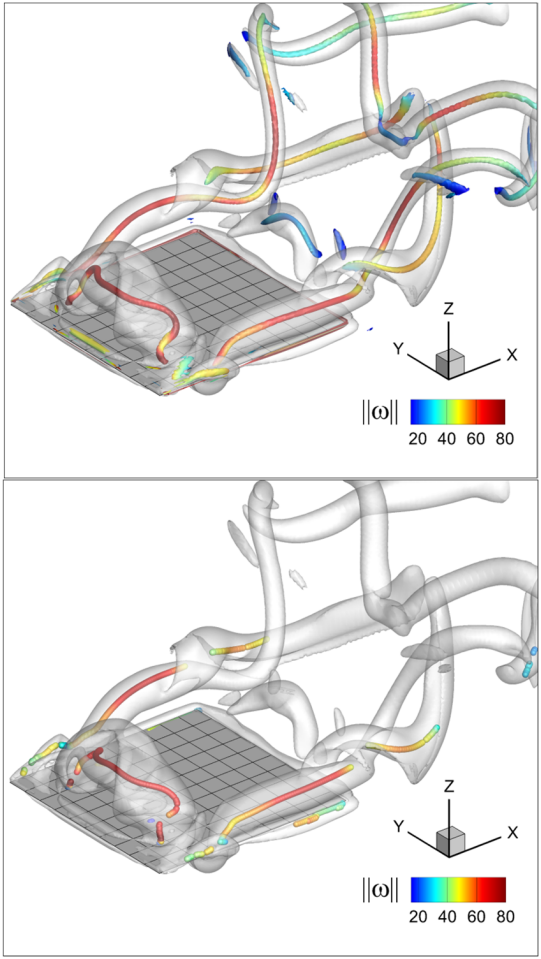

Fig. 8: Vortical structures for a heaving square plate. Isosurface of $Q = 60$ are shown in transparent grey, while isosurfaces of $D = 0.015$ show the vortex cores colored with vorticity magnitude (top). The bottom plate shows vortex core lines (thickened for clarity using lines with spheres instead of points) obtained with the method of Sujudi & Haimes [11].

### 3.6. *Containership propeller-rudder interaction*

In this case the propeller/rudder interaction flow of the KCS containership during a zigzag maneuver in shallow water is visualized [28]. Computations were performed with a hybrid RANS/LES turbulence modeling approach, and thus only the largest vortices are resolved. The instantaneous solution shown in Fig. 9 is taken from a simulation of KCS performing a 20/5 zigzag maneuver in shallow water with water depth to ship draft ratio $h/T = 1.2$. The ship is self-propelled at Froude number $\mathrm{Fr} = V_0 / \sqrt{L_0 g} = 0.095$, where $L_0$ and $V_0$ are the ship length and speed and used to non-dimensionalize all variables. The speed is controlled by an autopilot that varies the rotational velocity of the propeller. In a 20/5 maneuver test the rudder is turned in the opposite direction at the maximum rudder rate every time the ship heading checks +/− 5 degrees, resulting in a zigzag trajectory. In this case the propeller produces thrust mostly through pressure lift on the blades, resulting in high pressure gradients external to the vortex-induced pressure gradients.

As in the example in §3.4, the propeller produces strong hub and blade tip vortices, which mostly dominate the vortical structure of the flow. The rudder also produces vortices, and affects the propeller vortices through complex interaction. Figure 9 shows a view of the stern with the propeller/rudder system from below, with the propeller rotating clockwise looking from stern to bow. Isosurfaces of $Q = 15000$ can only display propeller/rudder vortices, since these are much stronger than other vortices generated by the flow, and have high dimensionless vorticity magnitude ($|\omega| > 200$). Isosurfaces of lower $Q$ produce large numbers of structures that obscure the picture. Isosurfaces of $D = 0.001$ show well the tip vortices far from the propeller where $Q$ fails, and also display more structures present in the flow, in particular the bilge vortex generated by the drift of the ship as it turns, and the weakened blade tip vortices downstream of the rudder. The bilge vortices, visible on the port side of the propeller shaft hub, are of special interest since they are ingested by the propeller, and they change strength and axial vorticity sign as the zigzag maneuver causes drift to starboard or port.

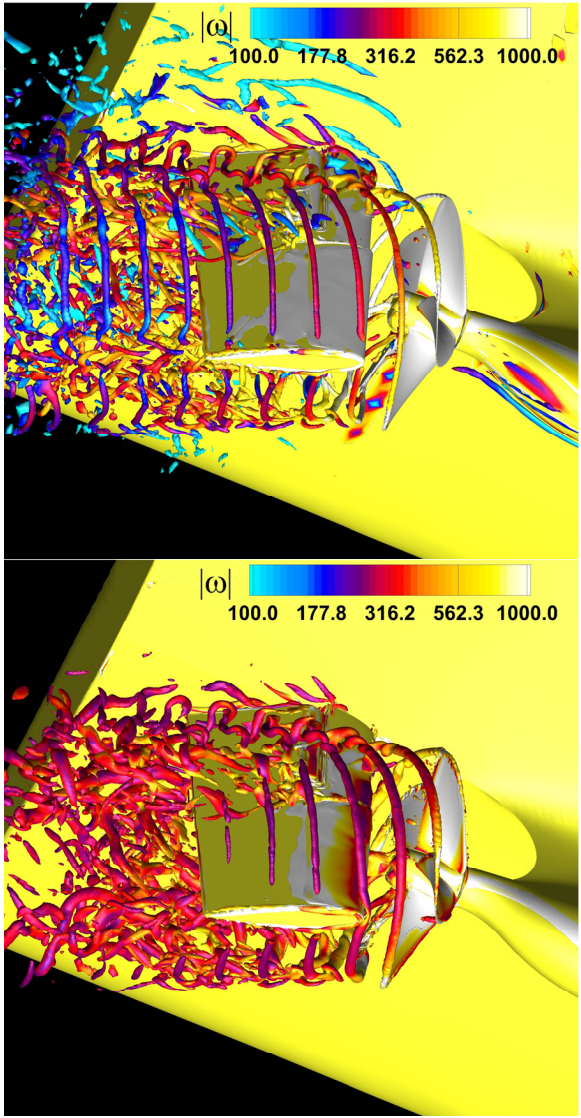

Fig. 9: Propeller-rudder interaction for containership in zigzag maneuver. Isosurfaces of $D = 0.001$ (top) and $Q = 15000$ (bottom) coloured with dimensionless vorticity magnitude.

## 4. Conclusion

We discussed a method to compute a function $D$ that provides the approximate distance to a vortex centerline for incompressible flows. The method assumes that the pressure gradient is mainly caused by the local vortex, as opposed to pressure gradients caused by other nearby vortices or flow boundaries. This effect of external pressure gradients poses limitations to the method as discussed in the paper, but in the many examples shown these limitations appear important only when weak vortices are affected by strong pressure gradients caused by the primary flow around a blunt body. Examples show that $D$ can effectively display vortex tubes around the core in most flows, and that can complement well other methods to visualize turbulent structures.

Depending on the flow configurations, properly simplified formulations of $D$ can be used for better visualization. However, in most the examples shown in this paper the simplest expression $D = |\nabla P|/\max(Q,\epsilon)$ works well. Suggested isosurfaces of $D$ of a few cell lengths at locations of interests with $\epsilon$ a couple of order of magnitude smaller than the lowest expected $Q$ produces tubes of approximate diameter $2D$.

## ACKNOWLEDGEMENTS

This work was partially supported by the US Office of Naval Research grants N00014-17-1-2082 and N00014-17-1-2293, Drs. Thomas Fu and Ki-Han Kim program officers.

## REFERENCES

[1] Chakraborty, P., Balachandar, S., and Adrian, R. J., 2005, "On the relationships between local vortex identification schemes," J. Fluid Mech., 535, pp. 189-214.

[2] Haller, G., 2005, "An objective definition of a vortex," J. Fluid Mech., 525, pp. 1-26.

[3] Jeong, J., and Hussain, F., 1995, "On the Identification of a Vortex," J. Fluid Mech., 285, pp. 69-94.

[4] Epps, B. P., 2017, "Review of Vortex Identification Methods," AIAA SciTec Forum.

[5] Volkov, K. N., Emel'yanov, V. N., Teterina, I. V., and Yakovchuk, M. S., 2017, "Visualization of vortical flows in computational fluid dynamics," Comput. Math. Math. Phys., 57(8), pp. 1360-1375.

[6] Hunt, J., Wray, A., and Moin, P., 1988, "Eddies, streams, and convergence zones in turbulent flows," Tech. Rep. CTR-S88. Center for Turbulence Research.

[7] Green, M. A., Rowley, C. W., and Haller, G., 2007, "Detection of Lagrangian coherent structures in three-dimensional turbulence," J. Fluid Mech., 572, pp. 111-120.

[8] Haller, G., 2015, "Lagrangian Coherent Structures," Annu. Rev. Fluid Mech., 47, pp. 137-162.

[9] Dubief, Y., and Delcayre, F., 2000, "On coherent-vortex identification in turbulence," J. Turbulence, 1, pp. 1-22.

[10] Zhou, J., Adrian, R. J., Balachandar, S., and Kendall, T. M., 1999, "Mechanisms for generating coherent packets of hairpin vortices in channel flow," J. Fluid Mech., 387, pp. 353-396.

[11] Sujudi, D., and Haimes, R., 1995, "Identification of swirling flow in 3-D vector fields," AIAA Paper No. AIAA 95-1715.

[12] Garth, C., Tricoche, X., Wiebel, A., and Joy, K., 2008, "On the role of domain-specific knowledge in the visualization of technical flows," Comp. Graphics Forum 27, 1007-1014.

[13] Roth, M., 2000, "Automatic extraction of vortex core lines and other line-type features for scientific visualization," PhD thesis, Swiss Federal Institute of Technology Zurich.

[14] Levy, Y., Degani, D., and Seginer, A., 1990, "Graphical Visualization of Vortical Flows by Means of Helicity," AIAA J., 28(8), pp. 1347-1352.

[15] Banks, D. C., and Singer, B. A., 1994, "Vortex Tubes in Turbulent Flows: Identification, Representation, Reconstruction," In Conference on Visualization, pp. 132-139.

[16] Kida, S., and Miura, H., 1998, "Identification and analysis of vortical structures," European J. Mech. B/Fluid, 17(4), pp. 471-488.

[17] Strawn, R. C., Kenwright, D. N., and Ahmad, J., 1999, "Computer Visualization of Vortex Wake Systems," AIAA J., 37(4), pp. 511-512.

[18] Gerz, T., Holzapfel, F., and Darracq, D., 2002, "Commercial aircraft wake vortices," Progr. Aerospace Sci., 38(3), pp. 181-208.

[19] Leweke, T., Le Dizes, S., and Williamson, C. H. K., 2016, "Dynamics and Instabilities of Vortex Pairs," Annu. Rev. Fluid Mech., 48, pp. 507-541.

[20] Bürger, K., Treib, M., Westermann, R., Werner, S., Lalescu, C., Szalay, A., Meneveau, C., and Eyink, G., 2012, "Vortices within vortices: hierarchical nature of vortex tubes in turbulence." arXiv:1210.3325v2 [physics.flu-dyn]

[21] Boedewadt, U. T., 1940, "Die Drehstromung uber festem Grund," Z. Angew. Math. Mech., 20, pp. 241-253.

[22] Schwiderski, E. W., and Lugt, H. J., 1964, "Rotating Flows of von Kármán and Bödewadt," Phys. Fluids, 7(6), pp. 867-875.

[23] Li, Y., Perlman, E., Wan, M. P., Yang, Y. K., Meneveau, C., Burns, R., Chen, S. Y., Szalay, A., and Eyink, G., 2008, "A public turbulence database cluster and applications to study Lagrangian evolution of velocity increments in turbulence," J. Turbulence, 9(31), pp. 1-29.

[24] Chase, N., and Carrica, P., 2013, "Submarine propeller computations and application to self-propulsion of DARPA Suboff," Ocean Eng., 60, pp. 68–80.

[25] Di Mascio, A., Muscari, R., and Dubbioso, G., 2014, "On the wake dynamics of a propeller operating in drift," J. Fluid Mech., 754, pp. 263-307.

[26] Leweke, T., Quaranta, H. U., Bolnot, H., Blanco-Rodriguez, F. J., and Le Dizes, S., 2014, "Long- and short-wave instabilities in helical vortices," Journal of Physics: Conference Series, 524, 012154

[27] Triantafyllou, M. S., Techet, A. H., and Hover, F. S., 2004, "Review of experimental work in biomimetic foils," IEEE J. Oceanic Eng., 29(3), pp. 585-594.

[28] Carrica, P., Mofidi, A., Eloot, K., and Delefortrie, G., 2016, "Direct simulation and experimental study of zigzag maneuver of KCS in shallow water," Ocean Eng., 112, pp. 117-133.